\documentclass[prd,twocolumn,groupedaddress,showpacs,nofootinbib]{revtex4}
\usepackage{graphicx}
\usepackage{dcolumn}
\usepackage{bm}
\usepackage{amssymb}
\usepackage{mathrsfs}
\usepackage{amsmath}
\usepackage{epsfig}
\usepackage[dvips]{color}
\usepackage{hhline}

\begin{document}

\title{ Large Lepton Asymmetry for Small Baryon Asymmetry and Warm Dark Matter }

\author{Pei-Hong Gu}
\email{peihong.gu@mpi-hd.mpg.de}

\affiliation{Max-Planck-Institut f\"{u}r Kernphysik, Saupfercheckweg
1, 69117 Heidelberg, Germany}

\begin{abstract}

We propose a resonant leptogenesis scenario in a $U(1)_{B-L}^{}$
gauge extension of the standard model to generate large lepton
asymmetries for cosmological baryon asymmetry and dark matter. After
$B-L$ number is spontaneously broken, inflaton can pick up a small
vacuum expectation value for the mass splits of three pairs of
quasi-degenerately heavy Majorana neutrinos and the masses of three
sterile neutrinos. With thermal mass effects of sphalerons, the
observed small baryon asymmetry can be converted from large lepton
asymmetries of individual flavors although total lepton asymmetry is
assumed zero. The mixing between sterile and active neutrinos is
elegantly suppressed by the heavy Majorana neutrinos. Before the
active neutrinos start their strong flavor conversions, the sterile
neutrinos as warm dark matter can be produced by resonant
active-sterile neutrino oscillations to reconcile X-ray and
Lyman-$\alpha$ bounds. Small neutrino masses are naturally realized
by seesaw contributions from the heavy Majorana neutrinos and the
sterile neutrinos.

\end{abstract}


\maketitle

\section{Introduction}

Sterile neutrinos can provide the dark matter relic density through
their oscillations with active neutrinos
\cite{dw1994,sf1999,als2007,ls2008,afp2001,dh2001,aft2001,abs2005}.
Because of the mixing with the active neutrinos, the sterile
neutrinos can decay at tree level and loop orders. In particular,
the decays of a sterile neutrino into an active neutrino and a
photon at one-loop order will produce a narrow line in the X-ray
background \cite{dh2001}. The X-ray constraints put an upper bound
on the sterile neutrino mass \cite{bnrst2006}. Furthermore, the
analysis on the Lyman-$\alpha$ data shows a low bound on the sterile
neutrino mass \cite{blrv2009}. The X-ray bound will be in conflict
with the Lyman-$\alpha$ bound if the sterile neutrino is produced by
the non-resonant oscillations between the active and sterile
neutrinos \cite{blrv2009}. In the presence of a large neutrino
asymmetry, the resonant active-sterile oscillations \cite{sf1999}
can reconcile the two bounds \cite{shaposhnikov2008}. Such a large
lepton asymmetry seems inconsistent with the small baryon asymmetry
of the universe because the lepton and baryon asymmetries are
usually enforced to be at a same order by sphalerons \cite{krs1985}.
This problem can be evaded in three ways: (1) the lepton asymmetry
is generated below the electroweak scale; (2) the sphaleron
transition doesn't work; (3) one type of lepton asymmetry is
canceled by an opposite lepton asymmetry of other flavors. These
possibilities have been proposed and studied in many works
\cite{ekm1990,as2005,linde1976,mmr1999,kty2002}.

On the other hand, observations of solar, atmospheric, reactor and
accelerator neutrino oscillations have established the massive and
mixing neutrinos \cite{stv2008}. The cosmological bound shows the
neutrino masses should be in the sub-eV range \cite{dunkley2008}.
The smallness of neutrino masses can be naturally explained in the
seesaw \cite{minkowski1977} extension of the standard model (SM).
The seesaw essence is to make the neutrino masses tiny via a
suppressed ratio of the electroweak scale over a high scale. Most
popular seesaw schemes need lepton number violation as the neutrinos
are assumed to be Majorana particles. In the seesaw context, the
cosmological baryon asymmetry can be understood via leptogenesis
\cite{fy1986,lpy1986,fps1995,pilaftsis1997,ms1998}, where a lepton
asymmetry is first produced and then is partially converted to a
baryon asymmetry through the sphaleron \cite{krs1985} transition. In
particular, the so-called resonant leptogenesis
\cite{fps1995,pilaftsis1997} models \cite{ab2003,fjn2003,gu2010}
with quasi-degenerately decaying particles can induce a CP asymmetry
of the order of unit. This allows the production of a large lepton
asymmetry if we don't take the observed small baryon asymmetry into
account.

In this paper, we show that by the resonant leptogenesis the large
neutrino asymmetries of individual flavors can be generated for the
production of the sterile neutrino dark matter. The total lepton
asymmetry is assumed zero so that the baryon asymmetry can arrive at
a correct value through the sphaleron processes with thermal mass
effects \cite{shaposhnikov1988}. We demonstrate this possibility in
a $U(1)_{B-L}^{}$ gauge extension of the SM. There are two singlet
and a doublet Higgs scalars with lepton numbers besides the SM one.
One Higgs singlet drives the spontaneous symmetry breaking of the
$U(1)_{B-L}^{}$. The other one is responsible for the chaotic
inflation \cite{linde1983}. After the $B-L$ number is spontaneously
broken, the inflaton can pick up a seesaw suppressed vacuum
expectation value (VEV) due to the small ratio of the $B-L$ breaking
scale over its heavy mass. In the fermion sector, we introduce three
types of SM singlet fermions including usual right-handed neutrinos.
Through the large VEV for the $B-L$ symmetry breaking, the
right-handed neutrinos can mix with one type of additional singlets,
which gets small Majorana masses from the VEV of the inflaton. So,
we can naturally have three pairs of quasi-degenerately heavy
Majorana neutrinos for the resonant leptogenesis. The other singlet
fermions obtain small masses through their Yukawa couplings with the
inflaton. They can play the role of the sterile neutrinos as their
mixing with the active neutrinos is seesaw suppressed. The resonant
active-sterile oscillation with neutrinos is induced by a positive
neutrino asymmetry while that with antineutrinos is induced by a
negative neutrino asymmetry. Although the total neutrino asymmetry
is vanishing, the resonant sterile-active oscillations are still
available as they occur much earlier than the beginning of the
strong flavor conversions of the active neutrinos. As for the
neutrino masses, they are generated in a seesaw scenario, where the
heavy Majorana neutrinos and the light sterile neutrinos give a
dominant and a negligible contribution, respectively.

\section{The model}

The field content of our model is summarized in Table \ref{fields},
where $(q_L^{},d_R^{},u_R^{})$ and $(\psi_L^{},l_R^{})$,
respectively, are the SM quarks and leptons, $\varphi$ is the SM
Higgs doublet, $\nu_R^{}$ denotes usual right-handed neutrinos,
$\xi_R^{}$ and $\zeta_R^{}$ are additional right-handed singlet
fermions, $\chi$ and $\sigma$ are two Higgs singlets, $\eta$ is a
new Higgs doublet. Note the new fermions $\nu_R^{}$, $\xi_R^{}$ and
$\zeta_R^{}$ all have three generations so that the model can be
free of gauge anomaly. The full Lagrangian should be
$SU(3)_{c}^{}\times SU(2)_{L}^{}\times U(1)_{Y}^{} \times
U(1)_{B-L}^{}$ invariant. For simplicity we only write down the part
relevant for our discussions,
\begin{eqnarray}
\label{lagrangian1} \mathcal{L} &\supset&
-y_\nu^{}\bar{\psi}_L^{}\varphi
\nu_R^{}-y_\xi^{}\bar{\psi}_L^{}\eta\xi_R^{}-f\chi\bar{\nu}_R^{c}\xi_R^{}-\frac{1}{2}h_\xi^{}\sigma^\ast_{}\bar{\xi}_{R}^{c}\xi_R^{} \nonumber\\
&&
-\frac{1}{2}h_\zeta^{}\sigma\bar{\zeta}_{R}^{c}\zeta_R^{}-\mu\bar{\zeta}_{R}^{c}\xi_R^{}
- \kappa\sigma \chi\varphi^\dagger_{}\eta  - \rho
\sigma^\ast_{}\chi^2_{}+\textrm{H.c.}\nonumber\\
&&-m_\sigma^2|\sigma|^2_{}-\lambda|\sigma|^2_{}\,.
\end{eqnarray}
Here the Yukawa couplings $h_\xi^{}$ and $h_\zeta^{}$ are symmetric.
In addition, the scalar parameters $\kappa$ and $\rho$ are rotated
to be real without loss of generality. Furthermore, the mass term
$m_\sigma^2$ is positive.

\begin{table}
\begin{center}
\caption{Quantum number assignments. Here $(q_L^{},d_R^{},u_R^{})$
and $(\psi_L^{},l_R^{})$, respectively, are the SM quarks and
leptons, $\varphi$ is the SM Higgs doublet, $\nu_R^{}$ denotes usual
right-handed neutrinos, $\xi_R^{}$ and $\zeta_R^{}$ are additional
right-handed singlet fermions, $\chi$ and $\sigma$ are two Higgs
singlets, $\eta$ is a new Higgs doublet. All of the fermions have
three families. } \label{fields} \vspace{2mm}
\begin{tabular}{|c|c|c|}  \hline &&\\[0.5mm]
~~~~~~~~~&~~~$SU(3)_{c}^{}\times SU(2)_{L}^{}\times U(1)_{Y}^{}$~~~& ~~~$U(1)_{B-L}^{}$~~~ \\
&&\\[-1.0mm]
\hline &&\\[-0.2mm]$q_L^{}$ &
$(\textbf{3},\textbf{2},+\frac{1}{6})$ &$+\frac{1}{3}$\\
&& \\[-2mm]
\hline  &&\\[-0.2mm]$d_R^{}$ &
$(\textbf{3},\textbf{1},-\frac{1}{3})$&$+\frac{1}{3}$\\
&& \\[-2mm]
\hline  &&\\ [-0.2mm] $u_R^{}$ &
$(\textbf{3},\textbf{1},+\frac{2}{3})$ & $+\frac{1}{3}$\\
&& \\[-2mm]
\hline  && \\[-0.2mm]$\psi_L^{}$ &
$(\textbf{1},\textbf{2},-\frac{1}{2})$ & $-1$\\
&& \\[-2mm]
\hline  &&\\[-0.2mm]$ l_R^{}$ &
$(\textbf{1},\textbf{1},-1)$ & $-1$\\
&& \\[-2mm]
\hline  &&\\[-0.2mm]$\varphi$ &
$(\textbf{1},\textbf{2},-\frac{1}{2})$ & $~~0$\\
&& \\[-2mm]
\hline  &&\\[-0.2mm]$ \nu_R^{}$ &
$(\textbf{1},\textbf{1},~~0)$ & $-1$\\
&& \\[-2mm]
\hline  &&\\[-0.2mm]$ \xi_R^{}$ &
$(\textbf{1},\textbf{1},~~0)$ & $+\frac{1}{2}$\\
&& \\[-2mm]
\hline  &&\\[-0.2mm]$ \zeta_R^{}$ &
$(\textbf{1},\textbf{1},~~0)$ & $-\frac{1}{2}$\\
&& \\[-2mm]
\hline  &&\\[-0.2mm]$\chi$ &
$(\textbf{1},\textbf{1},~~0)$ &$+\frac{1}{2}$\\
&& \\[-2mm]
\hline  &&\\[-0.2mm]$\sigma$ &
$(\textbf{1},\textbf{1},~~0)$ &$+1$ \\
&& \\[-2mm]
\hline  &&\\[-0.2mm]$\eta$ &
$(\textbf{1},\textbf{2},-\frac{1}{2})$ & $-\frac{3}{2}$\\
[2.0mm]\hline
\end{tabular}
\end{center}
\end{table}

\section{Heavy Majorana neutrinos}

The $U(1)_{B-L}^{}$ gauge symmetry will be spontaneously broken
after the Higgs singlet $\chi$ develops its VEV,
\begin{eqnarray}
\label{vev1} \langle\chi\rangle =v_\chi^{}\,.
\end{eqnarray}
The other Higgs singlet $\sigma$ will then pick up a small VEV,
\begin{eqnarray}
\label{vev2} \langle\sigma\rangle = v_\sigma^{}\simeq-\frac{\rho
v_\chi^2 }{m_{\sigma}^2}\,,
\end{eqnarray}
like the type-II seesaw \cite{mw1980} and its variation
\cite{gh2006}. At this stage, we can derive the following mass terms
from Eq. (\ref{lagrangian1}),
\begin{eqnarray}
\label{lagrangian2} \mathcal{L}
&\supset&-fv_\chi^{}\bar{\nu}_R^{c}\xi_R^{}-\frac{1}{2}h_\xi^{}v_\sigma^{}\bar{\xi}_{R}^{c}\xi_R^{}
-\frac{1}{2}h_\zeta^{}v_\sigma^{}\bar{\zeta}_{R}^{c}\zeta_R^{}
-\mu\bar{\zeta}_R^c\xi_R^{} \nonumber\\
&&+\textrm{H.c.}\,.
\end{eqnarray}

The influence of $\zeta_{R}^{}$ on $\nu_{R}^{}$ and $\xi_{R}^{}$ is
negligible for $\displaystyle{f v_\chi^{}\gg\mu}$. We thus focus on
the mass terms only involving $\nu_{R}^{}$ and $\xi_{R}^{}$. For
illustration, we choose the base where $f$ is diagonal and real,
i.e. $\displaystyle{f=\textrm{diag}\{f_1^{},f_2^{},f_3^{}\}}$. For
$\displaystyle{f v_\chi^{}\gg h_\xi^{}v_\sigma^{}}$, we can perform
the following rotation,
\begin{subequations}
\label{singlet}
\begin{eqnarray}
\label{singlet1}
\nu_{R_i^{}}^{}&\simeq&\frac{1}{\sqrt{2}}\left(N_{R_i^{}}^{+}-iN_{R_i^{}}^{-}\right)\,,\\
\label{singlet2}
\xi_{R_i^{}}^{}&\simeq&\frac{1}{\sqrt{2}}\left(N_{R_i^{}}^{+}+iN_{R_i^{}}^{-}\right)\,,
\end{eqnarray}
\end{subequations}
to obtain the diagonal masses,
\begin{eqnarray}
\label{lagrangian3} \mathcal{L} &\supset&
-\frac{1}{2}\bar{N}_R^{+c}m_{N^+_{}}^{}N_R^{+}-\frac{1}{2}\bar{N}_R^{-c}m_{N^-_{}}^{}N_R^{-}+
\textrm{H.c.}
\end{eqnarray}
with
\begin{subequations}
\label{mass}
\begin{eqnarray}
\label{mass1} m_{N^+_{i}}^{}&\simeq&f_{i}^{}v_\chi^{}
+ \frac{1}{2}h_{\xi_{ii}^{}}^{}v_\sigma^{}\,,\\
\label{mass2} m_{N^-_{i}}^{}&\simeq &f_{i}^{}v_\chi^{} -
\frac{1}{2}h_{\xi_{ii}^{}}^{}v_\sigma^{}\,.
\end{eqnarray}
\end{subequations}
It is then convenient to define the Majorana fermions,
\begin{subequations}
\label{majoranafermion}
\begin{eqnarray}
\label{majoranafermion1}
N_i^{+} &=& N_{R_i^{}}^{+}+ N_{R_i^{}}^{+c}\,,\\
\label{majoranafermion2} N_i^{-} &=& N_{R_i^{}}^{-}+
N_{R_i^{}}^{-c}\,.
\end{eqnarray}
\end{subequations}
Clearly, $N_i^+$ and $N_i^-$ are quasi-degenerate as their mass
split is much smaller than their masses.

We now derive the Yukawa couplings of the heavy Majorana fermions
$N_{}^{\pm}$. For convenience, we first define
\begin{eqnarray}
\label{doublets} \varphi=c\phi_{1}^{} +
s\phi_2^{}\,,~~\eta=-s\phi_{1}^{} + c\phi_2^{}
\end{eqnarray}
with
\begin{eqnarray}
\label{mixingangle} c\equiv \cos \vartheta\,,~~s\equiv \sin
\vartheta \,,~~\vartheta =\frac{1}{2}\arctan\frac{2\kappa
v_\chi^{}v_\sigma^{} }{\mu_\varphi^2-\mu_\eta^2}\,.
\end{eqnarray}
Here $\mu_\varphi^{2}$ and $\mu_\eta^{2}$ are the mass terms of
$\varphi$ and $\eta$, respectively. The mass terms of $\phi_1^{}$
and $\phi_2^{}$ would be
\begin{subequations}
\label{doublets2}
\begin{eqnarray}
\label{doublets21}
\mu_{\phi_1^{}}^2=\frac{1}{2}\left\{\mu_\varphi^{2}+\mu_\eta^{2}
-\left[\left(\mu_\varphi^{2}-\mu_\eta^{2}\right)^2_{}+4\kappa^2_{}
v_\chi^2 v_\sigma^2\right]^{\frac{1}{2}}_{}\right\}\,,~~&&\\
\label{doublets22}
\mu_{\phi_2^{}}^2=\frac{1}{2}\left\{\mu_\varphi^{2}+\mu_\eta^{2}
+\left[\left(\mu_\varphi^{2}-\mu_\eta^{2}\right)^2_{}+4\kappa^2_{}
v_\chi^2 v_\sigma^2\right]^{\frac{1}{2}}_{}\right\}\,.~~&&
\end{eqnarray}
\end{subequations}
At least one of $\mu_\varphi^{2}$ and $\mu_\eta^{2}$ should be
negative to guarantee the electroweak symmetry breaking, i.e.
\begin{eqnarray}
v=\sqrt{v_\varphi^{2}+v_\eta^{2}}\simeq
174\,\textrm{GeV}\quad\textrm{with}\quad\quad\quad\quad\quad\quad&&\nonumber\\
v_\varphi^{}=\langle\varphi\rangle>\frac{m_t^{}}{\sqrt{4\pi}}\simeq
\frac{171.2\,\textrm{GeV}}{\sqrt{4\pi}}\simeq
48\,\textrm{GeV}\,,\,v_\eta^{}=\langle\eta\rangle\,.&&
\end{eqnarray}
For example, we can take $\mu_{\varphi}^2<0<\mu_{\eta}^2$ and then
obtain
$\mu_{\phi_1^{}}^2<\mu_{\varphi}^2<0<\mu_{\eta}^2<\mu_{\phi_2^{}}^2$.
By inserting Eqs. (\ref{singlet}), (\ref{majoranafermion}) and
(\ref{doublets}) to the first and second terms of Eq.
(\ref{lagrangian1}), we eventually obtain
\begin{eqnarray}
\label{lagrangian4} \mathcal{L}
\supset-\left(y_\pm^{a}\right)_{\alpha
i}^{}\bar{\psi}_{L_\alpha^{}}^{}\phi_a^{} N_{i}^{\pm}+\textrm{H.c.}
\end{eqnarray}
with
\begin{subequations}
\label{yukawa1}
\begin{eqnarray}
\label{yukawa11} y^1_+=\frac{1}{\sqrt{2}}\left(c y_\nu^{} -
sy_\xi^{}\right)\,,~~y^1_-=-\frac{i}{\sqrt{2}}\left(c y_\nu^{}
+s y_\xi^{}\right)\,,&&\\
\label{yukawa12} y^2_+=\frac{1}{\sqrt{2}}\left(s y_\nu^{} + c
y_\xi^{}\right)\,,~~y^2_-=-\frac{i}{\sqrt{2}}\left(s y_\nu^{} - c
y_\xi^{}\right)\,.&&
\end{eqnarray}
\end{subequations}
Clearly, the heavy Majorana fermions $N^{\pm}_{}$ play the same role
with the heavy Majorana neutrinos in the usual seesaw model. We thus
refer to $N^{\pm}_{}$ as the heavy Majorana neutrinos.

\section{Resonant leptogenesis}

From Eq. (\ref{mass}), it is straightforward to see the heavy
Majorana neutrinos $N^{+}_{i}$ and $N^{-}_{i}$ have a very small
mass split compared to their masses. This means the Yukawa
interaction (\ref{lagrangian4}) is probably ready for the resonant
leptogenesis \cite{fps1995,pilaftsis1997} if other conditions are
satisfied. Because of the special texture of the Yukawa couplings
(\ref{yukawa1}), the decays of $N^{\pm}_{i}$ can not generate a
nonzero lepton asymmetry if the two Higgs doublets $\phi_{1,2}^{}$
both appear in the final states. This could be easily understood in
the base with $(\nu_{R_i^{}}^{},\xi_{R_i^{}}^{})$ and
$(\varphi,\eta)$. Since we have ignored the small Majorana mass term
$h_\xi^{} v_\sigma^{}$ for giving the rotation (\ref{singlet}) and
then the Yukawa couplings (\ref{yukawa1}), both $\nu_{R_i^{}}^{}$
and $\xi_{R_i^{}}^{}$ only has one decay channel, i.e.
$\nu_{R_i^{}}^{} \rightarrow \psi_{L}^{} +
\varphi^\ast_{}\,,~\xi_{R_i^{}}^{} \rightarrow\psi_{L}^{} +
\eta^\ast_{}$. In the presence of the $\varphi-\eta$ mixing
(\ref{doublets}), we further have the decay channels,
$\nu_{R_i^{}}^{} \rightarrow\psi_{L}^{} +
\eta^\ast_{}\,,~\xi_{R_i^{}}^{} \rightarrow \psi_{L}^{} +
\varphi^\ast_{}$. Clearly, the decays into
$\psi_{L}^{}\varphi^\ast_{}\,,~\psi_{L}^c\varphi$ and those into
$\psi_{L}^{}\eta^\ast_{}\,,~\psi_{L}^c\eta$ will produce an equal
but opposite lepton asymmetry stored in the lepton doublets if the
CP is not conserved. For a successful leptogensis, we thus need one
of $\phi_{1,2}^{}$ to be heavier than the lightest pair of
$N^{\pm}_{1,2,3}$. For example, we choose $\phi_2^{}$ to be heavier
than $N^{\pm}_{1}$. The other heavy Majorana neutrinos
$N^{\pm}_{2,3}$, which are assumed much heavier than $N^{\pm}_{1}$,
have flexibilities to be heavier or lighter than $\phi_2^{}$.
Therefore a final lepton asymmetry would be produced by the two-body
decays of $N^{\pm}_{1}$, i.e.
\begin{eqnarray}
N_1^\pm &\rightarrow& \psi_{L_\alpha^{}}^{} + \phi_1^{\ast}\,,\quad
\psi_{L_\alpha^{}}^{c} + \phi_1^{}\,.
\end{eqnarray}
Following the standard method \cite{pilaftsis1997} of the resonant
leptogenesis, we can calculate the electron, muon and tau types of
lepton asymmetries from the decays of per $N_1^\pm$,
\begin{eqnarray}
\label{cpasymmetry}
&&\varepsilon_{N_1^\pm}^{\nu_\alpha^{}}=\varepsilon_{N_1^\pm}^{l_{L_\alpha^{}}^{}}=\frac{1}{2}\varepsilon_{N_1^\pm}^{L_\alpha^{}}\nonumber\\
&\simeq&\frac{1}{2} \frac{\Gamma(N_1^\pm \rightarrow
\psi_{L_\alpha^{}}^{} + \phi_1^{\ast})-\Gamma(N_1^\pm \rightarrow
\psi_{L_\alpha^{}}^{c} +
\phi_1^{})}{\sum_\alpha^{}\left[\Gamma(N_1^\pm \rightarrow
\psi_{L_\alpha^{}}^{} + \phi_1^{\ast})+\Gamma(N_1^\pm \rightarrow
\psi_{L_\alpha^{}}^{c} +
\phi_1^{})\right]}\nonumber\\
\vspace{10mm} &\simeq&\frac{sc}{16\pi
A_{N_1^\pm}^{}}\left\{\left[c^2_{}\left|y_{\nu_{\alpha
1}^{}}^{}\right|^2_{}-s^2_{}\left|y_{\xi_{\alpha
1}^{}}^{}\right|^2_{}\right]\textrm{Im}\left[\left(y_{\xi}^{\dagger}
y_{\nu}^{}\right)_{11}^{}\right]\right.\nonumber\\
&&\left.+\left[c^2_{}\left(y_\nu^\dagger
y_\nu^{}\right)_{11}^{}-s^2_{}\left(y_\xi^\dagger
y_\xi^{}\right)_{11}^{}\right]\textrm{Im}\left(y_{\xi_{\alpha
1}^{}}^\ast y_{\nu_{\alpha 1}^{}}^{}\right)\right\}\nonumber\\
&&\times \frac{r_{N_1^{}}^{}}{r_{N_1^{}}^{2}
+\frac{1}{64\pi^2_{}}A_{N_1^\mp}^{2}}
\end{eqnarray}
with
\begin{eqnarray}
A_{N_1^\pm}^{}&=&\frac{1}{2}\left\{s^2_{}\left(y_\xi^\dagger
y_\xi^{}\right)_{11}^{}+c^2_{}\left(y_\nu^\dagger
y_\nu^{}\right)_{11}^{}\right.\nonumber \\
&&\left.\mp sc\left[\left(y_\xi^\dagger
y_\nu^{}\right)_{11}^{}+\left(y_\nu^\dagger
y_\xi^{}\right)_{11}^{}\right]\right\}\nonumber\\
&=&\frac{1}{2}\sum_{\alpha}^{}\left(s^2_{}\left|y_{\xi_{\alpha
1}^{}}^{}\right|^2_{}+ c^2_{}\left|y_{\nu_{\alpha
1}^{}}^{}\right|^2_{}\right.\nonumber\\
&&\left.\mp 2sc\left|y_{\xi_{\alpha
1}^{}}^{}\right|\left|y_{\nu_{\alpha
1}^{}}^{}\right|\cos\delta_{\alpha 1}^{}\right)\,.
\end{eqnarray}
Here the parameter $r_{N_i^{}}^{}$ describes the mass split between
$N_i^{+}$ and $N_i^{-}$,
\begin{eqnarray}
\label{ratio}
r_{N_i^{}}^{}&=&\frac{m_{N_i^+}^2-m_{N_i^-}^2}{m_{N_i^+}^{}m_{N_i^-}^{}}=\frac{2h_{\xi_{ii}^{}}^{}v_\sigma^{}}{f_i^{}v_\chi^{}}\,.
\end{eqnarray}

The baryon and lepton asymmetries are determined by the $B-L$
asymmetry in the presence of sphalerons \cite{krs1985}. In the
present model, we have \cite{shaposhnikov1988}
\begin{eqnarray}
B=\frac{28}{79}\left(B-\sum_\alpha^{}L_\alpha^{}\right)=-\frac{28}{79}\sum_\alpha^{}L_\alpha^{}=-\frac{56}{79}\sum_\alpha^{}L_{\nu_\alpha^{}}^{}\,.&&\nonumber\\
&&
\end{eqnarray}
The masses and interactions will give corrections to the above
formula \cite{shaposhnikov1988},
\begin{eqnarray}
\Delta
B&=&-A\frac{6}{13\pi^2_{}}\sum_{\alpha}^{}\frac{\bar{m}^2_{\alpha}(T)}{T^2_{}}\left(L_\alpha^{}-\frac{1}{3}B\right)\nonumber\\
&=&-A\frac{6}{13\pi^2_{}}\sum_{\alpha}^{}\frac{\bar{m}^2_{\alpha}(T)}{T^2_{}}L_\alpha^{}\nonumber\\
&=&-A\frac{12}{13\pi^2_{}}\sum_{\alpha}^{}\frac{\bar{m}^2_{\alpha}(T)}{T^2_{}}L_{\nu_\alpha^{}}^{}\,.
\end{eqnarray}
Here the coefficient $A\simeq 1$ \cite{shaposhnikov1988}. In the
case that the sphaleron is still active after a weakly first-order
electroweak phase transition \cite{rt1999}, one finds
\cite{shaposhnikov1988}
\begin{eqnarray}
\frac{\bar{m}^2_{\alpha}(T)}{T^2_{}}=\frac{1}{6}f_\alpha^2+\frac{1}{3}f_\alpha^2\left[\frac{v(T)}{T}\right]^2_{}\leq
\frac{1}{2}f_\alpha^2
\end{eqnarray}
for $B-\Sigma_\alpha^{}L_\alpha^{}=0$. Here $f_\alpha^{}$ denotes
the Yukawa couplings of the electron, muon and tau to the SM Higgs.

With the thermal mass effects of the sphaleron processes, it is
possible to generate a small observed baryon asymmetry from large
lepton asymmetries of individual flavors \cite{mmr1999}. For this
purpose, we take
\begin{eqnarray}
\label{cpassumption} \textrm{Im}\left[\left(y_{\xi}^{\dagger}
y_{\nu}^{}\right)_{11}^{}\right]=\sum_\alpha^{}\left|y_{\xi_{\alpha
1}^{}}^{}\right|\left| y_{\nu_{\alpha
1}^{}}^{}\right|\sin\delta_{\alpha 1}^{}=0
\end{eqnarray}
to give a zero total lepton asymmetry. Under this assumption, the CP
asymmetry (\ref{cpasymmetry}) can be simplified by
\begin{eqnarray}
\label{cpasymmetry2}
&&\varepsilon_{N_1^\pm}^{\nu_\alpha^{}}=\varepsilon_{N_1^\pm}^{l_{L_\alpha^{}}^{}}=\frac{1}{2}\varepsilon_{N_1^\pm}^{L_\alpha^{}}\nonumber\\
&=&\frac{sc}{16\pi A_{N_1^\pm}^{}}\left[c^2_{}\left(y_\nu^\dagger
y_\nu^{}\right)_{11}^{}-s^2_{}\left(y_\xi^\dagger
y_\xi^{}\right)_{11}^{}\right]\sin\delta_{\alpha
1}^{}\nonumber\\
&&\times\left|y_{\xi_{\alpha 1}^{}}^{}\right|\left| y_{\nu_{\alpha
1}^{}}^{}\right|\frac{r_{N_1^{}}^{}}{r_{N_1^{}}^{2}
+\frac{1}{64\pi^2_{}}A_{N_1^\mp}^{2}}\,.
\end{eqnarray}
In the weak washout region where the out-of-equilibrium condition is
described by the quantity,
\begin{eqnarray}
K_{N_1^{\pm}}^{}=\frac{\Gamma_{N_1^\pm}^{}}{2H(T)}\left|_{T=m_{N_1^\pm}^{}}^{}\right.<1
\end{eqnarray}
with the decay width,
\begin{eqnarray}
\Gamma_{N_1^\pm}^{}&\simeq& \sum_\alpha^{}\left[\Gamma(N_1^\pm
\rightarrow \psi_{L_\alpha^{}}^{}+ \phi_1^{\ast}) \right.\nonumber\\
&&\left.+\Gamma(N_1^\pm \rightarrow \psi_{L_\alpha^{}}^{c} +
\phi_1^{})\right]\nonumber\\
&=&\frac{1}{8\pi}\left(y_{\pm}^{1^{}\dagger_{}}y_{\pm}^{1^{}_{}}\right)_{11}^{}m_{N_1^\pm}^{}=\frac{1}{8\pi}A_{N_1^\pm}^{}m_{N_1^\pm}^{}
\end{eqnarray}
and the Hubble constant
\begin{eqnarray}
\label{hubble}
H(T)=\left(\frac{8\pi^{3}_{}g_{\ast}^{}}{90}\right)^{\frac{1}{2}}_{}
\frac{T^{2}_{}}{M_{\textrm{Pl}}^{}}\,,
\end{eqnarray}
the final neutrino asymmetry can be approximately given by
\cite{kt1990}
\begin{eqnarray}
\label{neutrinoasymmetry} \eta_{\nu_\alpha^{}}^{}=
\frac{n_{\nu_\alpha^{}}^{}-n_{\bar{\nu}_\alpha^{}}^{}}{s}\simeq\frac{\varepsilon_{N_1^\pm}^{\nu_\alpha^{}}
}{g_\ast^{}}\,.
\end{eqnarray}
Here $g_{\ast}^{}\simeq 112$ is the relativistic degrees of freedom
(the SM fields plus $\zeta_{R_{i,2,3}^{}}^{}$) while
$M_{\textrm{Pl}}^{}\simeq 1.22\times 10^{19}_{}\,\textrm{GeV}$ is
the Planck mass. The final baryon asymmetry then should be
\begin{eqnarray}
\eta_B^{}=\frac{n_B^{}}{s}
\simeq-\frac{6}{13\pi^2_{}}\sum_\alpha^{}f_\alpha^2
\eta_{\nu_\alpha^{}}^{}\,.
\end{eqnarray}
Note the lepton number violating processes mediated by the heavier
$N_{2,3}^{\pm}$ should be decoupled before the leptogenesis epoch
$T=M_{N_1^{\pm}}^{}$ to give the solution (\ref{neutrinoasymmetry}).

The neutrino asymmetry $\eta_{\nu_\alpha^{}}^{}$ is related to the
neutrino chemical potential $\mu_{\nu_\alpha^{}}^{}$ by
\begin{eqnarray}
\eta_{\nu_\alpha^{}}^{}&=&\frac{15}{4\pi^2_{}g_{\ast
S}^{}(T)}\xi_{\nu_\alpha^{}}^{}+\mathcal{O}\left(\xi_{\nu_\alpha^{}}^{3}\right)\nonumber\\
&\simeq&\frac{15}{64\pi^2_{}}\xi_{\nu_\alpha^{}}^{}\quad\textrm{with}\quad
\xi_{\nu_\alpha^{}}^{}=\frac{\mu_{\nu_\alpha^{}}^{}}{T_{\nu_\alpha^{}}^{}}
\,.
\end{eqnarray}
Here we have taken $g_{\ast S}^{}(T)=16$ (photon, three neutrinos,
electron, positron plus three sterile neutrinos). The electron
neutrino asymmetry is tightly constrained by Primordial Big-Bang
Nucleosynthesis (BBN) \cite{sg2005},
\begin{eqnarray}
\label{bbn} \eta_{\nu_e^{}}^{}\in(-0.9,1.7)\times 10^{-3}_{}\quad
\textrm{for}\quad \xi_{\nu_e^{}}^{}\in(-0.04,0.07)\,.
\end{eqnarray}
The above bound also applies to the muon and tau neutrino
asymmetries because the neutrino oscillations will begin at
$10\,\textrm{MeV}$ to achieve strong flavor conversions before BBN
\cite{ls2000}. On the other hand, the five-year observations of the
WMAP collaboration precisely measured the baryon asymmetry as
\cite{dunkley2008}
\begin{eqnarray}
\label{cmb}
\eta_{B}^{}&=&\frac{1}{7.04}\times (6.225\pm0.170)\times 10^{-10}_{}\nonumber\\
&=&(0.884\pm0.024)\times 10^{-10}_{}\,.
\end{eqnarray}

\section{Sterile and active neutrinos}

During the evolution of the universe, the electroweak symmetry
breaking will happen when the Higgs doublet
\begin{eqnarray}
H =\frac{v_\varphi^{}}{v}\varphi+\frac{v_\eta^{}}{v}\eta\,
\end{eqnarray}
develops its VEV. The mass terms (\ref{lagrangian2}) then should be
extended to
\begin{eqnarray}
\label{lagrangian5} \mathcal{L}
&\supset&-y_\nu^{}v_\varphi^{}\bar{\nu}_L^{}\nu_R^{}-y_\xi^{}v_\eta^{}\bar{\nu}_L^{}\xi_R^{}-fv_\chi^{}\bar{\nu}_R^{c}\xi_R^{}
-\frac{1}{2}h_\xi^{}v_\sigma^{}\bar{\xi}_{R}^{c}\xi_R^{}\nonumber\\
&& -\frac{1}{2}h_\zeta^{}v_\sigma^{}\bar{\zeta}_{R}^{c}\zeta_R^{}
-\mu\bar{\zeta}_R^c\xi_R^{} +\textrm{H.c.}\,.
\end{eqnarray}
For convenience, we rewrite the above mass terms to be
\begin{eqnarray}
\mathcal{L} \supset-\frac{1}{2}(\bar{\nu}_L^{},\bar{\zeta}_R^c,
\bar{\nu}_R^c, \bar{\xi}_R^c) \mathcal{M}
(\nu_L^c,\zeta_R^{},\nu_R^{},\xi_R^{})^T_{}+\textrm{H.c.}\,,
\end{eqnarray}
where the mass matrix $\mathcal{M}$ is defined by
\begin{eqnarray}
\label{massmatrix1} \mathcal{M} =\left( \begin{array}{cc|cc}
~0~& ~0~ & ~y_\nu^{}v_\varphi^{}~ & ~y_\xi^{}v_\eta^{}~ \\
[2mm]
0  &h_\zeta^{}v_\sigma^{}& 0 & \mu \\
[3mm]\hline~&~&~&~\\
y_\nu^T v_\varphi^{}& 0 & 0 & f v_\chi^{}\\
[2mm] y_\xi^T v_\eta^{} & \mu_{}^T & f^T_{}v_\chi^{}& h_\xi^{}
v_\sigma^{}
\end{array}\right)\,.
\end{eqnarray}

For $fv_\chi^{}\gg
y_\nu^{}v_\varphi^{},y_\xi^{}v_\eta^{},h_\zeta^{}v_\sigma^{}~\textrm{and}~\mu^{}$,
we can make use of the seesaw formula to diagonalize the above mass
matrix $\mathcal{M}$ in two blocks,
\begin{eqnarray}
\label{lagrangian6} \mathcal{L}
&\supset&-\frac{1}{2}(\bar{\nu}_R^c,\bar{\xi}_R^c) \mathfrak{M}
(\nu_R^{},\xi_R^{})^T_{}-\frac{1}{2}(\bar{\nu}_L^{},\bar{\zeta}_R^c)
\mathfrak{m} (\nu_L^c,\zeta_R^{})^T_{}\nonumber\\
&&+\textrm{H.c.}\,,
\end{eqnarray}
where the mass matrices $\mathfrak{M}$ and $\mathfrak{m}$ are given
by
\vspace{9mm}
\begin{widetext}
\begin{eqnarray}
\label{matrix} \mathfrak{M}= \left(
\begin{array}{c|c}
0~~&~~ f v_\chi^{}  \\
[5mm]\hline
~&~\\
 f^T_{} v_\chi^{}~~&~~h_\xi^{}
v_\sigma^{}
\end{array}\right)\,,\quad
\mathfrak{m}= \left(
\begin{array}{c|c}
\displaystyle{y_\nu^{}\frac{1}{f^T_{}}h_\xi^{}\frac{1}{f}y_\nu^T\frac{v_\sigma^{}v_\varphi^2}{v_\chi^2}
-(y_\nu^{}\frac{1}{f^{T}_{}}y_\xi^T+y_\xi^{}\frac{1}{f}y_\nu^T)\frac{v_\varphi^{}v_\eta^{}}{v_\chi^{}}}~~
& ~~-\displaystyle{y_\nu^{}\frac{1}{f^T_{}}\mu^T_{}\frac{v_\varphi^{}}{v_\chi^{}}} \\
[5mm]\hline
~&~\\
-\displaystyle{\mu\frac{1}{f}y_\nu^T\frac{v_\varphi^{}}{v_\chi^{}}}~~&~~
h_\zeta^{}v_\sigma^{}
\end{array}\right)\,.
\end{eqnarray}
\end{widetext}
Clearly, the heavy mass matrix $\mathfrak{M}$ will give us the
quasi-degenerately heavy Majorana neutrinos $N^\pm_{}$. As for the
light mass matrix $\mathfrak{m}$, it can also accommodate the seesaw
if its diagonal element $h_\zeta^{}v_\sigma^{}$ is much bigger than
other elements. In this seesaw scenario, the neutrino masses should
be
\begin{eqnarray}
\label{lagrangian8} \mathcal{L}\supset-\frac{1}{2}\bar{\nu}_L^{}
m_\nu^{} \nu_L^c+\textrm{H.c.}=-\frac{1}{2}\bar{\nu}_L^{}
(m_\nu^{N}+m_\nu^{S}) \nu_L^c+\textrm{H.c.}&&\nonumber\\
&&
\end{eqnarray}
with
\begin{subequations}
\begin{eqnarray}
\label{numass1}
m_\nu^{N}&=&y_\nu^{}\frac{1}{f^T_{}}h_\xi^{}\frac{1}{f}y_\nu^T\frac{v_\sigma^{}v_\varphi^2}{v_\chi^2} \nonumber\\
&&-(y_\nu^{}\frac{1}{f^{T}_{}}y_\xi^T+y_\xi^{}\frac{1}{f}y_\nu^T)\frac{v_\varphi^{}v_\eta^{}}{v_\chi^{}}\,,\\
\label{numass2}
m_\nu^{S}&=&y_\nu^{}\frac{1}{f^T_{}}\mu^T_{}\frac{1}{h_\zeta^{}}\mu\frac{1}{f}y_\nu^T\frac{v_\varphi^2}{v_\sigma^{}v_\chi^2}\,.
\end{eqnarray}
\end{subequations}

The above neutrino mass matrices are expected to explain the
neutrino oscillation experiments \cite{stv2008},
\begin{eqnarray}
&\Delta m_{21}^2 =7.65^{+0.23}_{-0.20} \times
10^{-5}_{}\,\textrm{eV}^2_{}\,,~\sin\theta_{12}^2 = 0.304^{+0.022}_{-0.016}\,, &\nonumber\\
\vspace{10mm} &|\Delta m_{31}^2 |= 2.4^{+0.12}_{-0.11}\times
10^{-3}_{}\,\textrm{eV}^2_{}\,,~\sin\theta_{23}^2 =
0.50^{+0.07}_{-0.06} \,,~~&\nonumber\\
\vspace{8mm} &\sin\theta_{13}^2 =0.01^{+0.016}_{-0.011}\,.&
\end{eqnarray}
On the other hand, $\zeta_R^{}$ plays the role of the sterile
neutrinos, i.e.
\begin{eqnarray}
\label{lagrangian7}
\mathcal{L}\supset-\frac{1}{2}h_\zeta^{}v_\sigma^{}\bar{\zeta}_R^c
\zeta_R^{}+\textrm{H.c.}=-\frac{1}{2}m_S^{}\bar{S} S\,,
\end{eqnarray}
where we have rotated
$h_\zeta^{}=\textrm{diag}\{h_{\zeta_1^{}}^{},h_{\zeta_2^{}}^{},h_{\zeta_3^{}}^{}\}$
and then defined
\begin{eqnarray}
S_i^{}=\zeta_{R_i^{}}^{}+\zeta_{R_i^{}}^c\quad\textrm{with}\quad
m_{S_i^{}}^{}=h_{\zeta_i^{}}^{}v_\sigma^{}\,.
\end{eqnarray}
The active-sterile mixing angle is
\begin{eqnarray}
\label{asmixing} \theta_{\alpha
i}^2=\frac{\left|\displaystyle{y_{\nu_{\alpha
j}}^{}\frac{1}{f_j^{}}\mu^{}_{ij}\frac{
v_\varphi^{}}{v_\chi^{}}}\right|^2_{}}{m_{S_i^{}}^2}~~\textrm{and}~~\sin^2_{}2\theta_i^{}=4\sum_{\alpha}^{}\theta_{\alpha
i}^2\,.
\end{eqnarray}

The sterile neutrinos can be produced through the active-sterile
neutrino oscillations. Specifically, the $\nu_\alpha^{}\rightarrow
S_i^{}$ or $\bar{\nu}_\alpha^{}\rightarrow S_i^{}$ oscillation is
determined by the sterile neutrino mass $m_{S_i^{}}^{}$, the
active-sterile mixing angle $\theta_{\alpha i}^{}$ and the neutrino
asymmetry $\eta_{\nu_\alpha^{}}^{}$. In order to compare with other
works, we define the following function \cite{ls2008,nr1988},
\begin{eqnarray}
\label{lafunction} c_{\alpha\alpha}^{}&=&\sqrt{2}G_{\textrm{F}}^{}[2
\eta_{\nu_\alpha^{}}^{}+\sum_{\beta\neq
\alpha}^{}\eta_{\nu_\beta^{}}^{}+(1+2\sin^2_{}\theta_W^{})\eta_{l_{L_\alpha^{}}^{}}^{}\nonumber\\
&&-(1-2\sin^2_{}\theta_W^{})\sum_{\beta\neq
\alpha}\eta_{l_{L_\beta^{}}^{}}^{}+2\sin^2_{}\theta_W^{}\sum_\beta^{} \eta_{l_{R_\beta^{}}^{}}^{}]\nonumber\\
&=&3\sqrt{2}G_{\textrm{F}}^{}\eta_{\nu_\alpha^{}}^{}
\end{eqnarray}
Here we have taken $\eta_{\nu_\alpha^{}}^{}=\eta_{l_\alpha^{}}^{}$,
$\sum_{\alpha}^{}\eta_{\nu_\alpha^{}}^{}=0$ and $\eta_{l_R^{}}^{}=0$
into account. A positive neutrino asymmetry
$\eta_{\nu_\alpha^{}}^{}>0$ can induce a MSW \cite{wolfenstein1978}
resonant behavior in the neutrino oscillation
$\nu_\alpha^{}\rightarrow S_i^{}$ whereas a negative one
$\eta_{\nu_\alpha^{}}^{}<0$ can enhance the antineutrino oscillation
$\bar{\nu}_\alpha^{}\rightarrow S_i^{}$ \cite{nr1988,sf1999}. It has
been pointed out that a large neutrino asymmetry can reconcile the
contradiction between the X-ray and Lyman-$\alpha$ bounds. For
example, in the work \cite{blrv2009-2,brs2009} where the lepton
asymmetry is flavor blind, i.e.
$\eta_{\nu_\alpha^{}}^{}=\eta_{l_{L_\alpha^{}}^{}}^{}=\eta_{l_{R_\alpha^{}}^{}}^{}=\eta_{\nu_e^{}}^{}$,
the authors show with the parameter choice,
\begin{eqnarray}
m_{S_i^{}}^{}\simeq
8\,\textrm{keV}\,,~~\sin^2_{}2\theta_{i}^{}\simeq
10^{-12}_{}\,,~~\eta_{\nu_e^{}}^{}=7\times10^{-5}_{}\,,
\end{eqnarray}
the resonant $\nu_\alpha^{}\rightarrow S_i^{}$ oscillation can
produce adequate $S_i^{}$ for the dark matter relic density. We
should keep in mind that the resonant active-sterile conversion
happens at a temperature of the order of $100\,\textrm{MeV}$ for the
sterile neutrino mass $m_{S_i^{}}^{}$ being a few keV
\cite{als2007,ls2008}. This means the electron neutrino asymmetry
$\eta_{\nu_e^{}}^{}$ and the opposite muon and tau neutrino
asymmetries $\eta_{\nu_{\mu,\tau}^{}}^{}$ can survive for the
resonant enhancement either in the neutrino oscillations or in the
antineutrino oscillations since the active neutrino oscillations
driven by $\Delta m_{31}^2$ and $\Delta m_{21}^2$ begin at lower
temperatures $T\sim10\,\textrm{MeV}$ and $T\sim3\,\textrm{MeV}$,
respectively.

\section{Parameter choice}

We now take reasonable choice of parameters to give the observed
small cosmological baryon asymmetry from large lepton asymmetries
for the dark matter production. Firstly, the heavy Majorana
neutrinos $N_{i}^\pm$ couple to the gauge boson $Z_{B-L}^{}$
associated with the $U(1)_{B-L}^{}$ symmetry. For
$m_{N_{1}^\pm}^{}=\mathcal{O}\left(10^5_{}\,\textrm{GeV}\right)$,
the Higgs singlet $\chi$ should have a VEV bigger than
$\mathcal{O}\left(10^{8}_{}\,\textrm{GeV}\right)$ to guarantee the
departure from equilibrium of $N_{1}^\pm$ at $T\simeq
M_{N_1^\pm}^{}$, unless the corresponding gauge coupling
$g_{B-L}^{}$ is fine tuned very small \cite{ky2005}. So, we take
\begin{eqnarray}
v_\chi^{}=\mathcal{O}(10^8_{}\,\textrm{GeV})
\end{eqnarray}
and then
\begin{eqnarray}
m_{N_{1}^\pm}^{}~&=&10^5_{}\,\textrm{GeV}\quad
\textrm{for}\quad f_1^{}~~=\mathcal{O}(10^{-3}_{})\,,\nonumber\\
m_{N_{2,3}^\pm}^{}&=&10^6_{}\,\textrm{GeV}\quad \textrm{for}\quad
f_{2,3}^{}=\mathcal{O}(10^{-2}_{})\,.
\end{eqnarray}
Secondly, the Higgs singlet $\sigma$ is expected to realize the nice
picture of the chaotic inflation \cite{linde1983}. This suggests
\cite{lr1998}
\begin{eqnarray}
m_\sigma^{}=\mathcal{O}(10^{13}_{}\,\textrm{GeV})\,,\quad
\lambda=\mathcal{O}(10^{-13}_{})\,.
\end{eqnarray}
We then conveniently set
\begin{eqnarray}
\rho=v_\chi^{}
\end{eqnarray}
in Eq. (\ref{vev2}) to induce a small VEV,
\begin{eqnarray}
\label{vev3} v_\sigma^{}=\mathcal{O}(10\,\textrm{MeV})\,.
\end{eqnarray}
In consequence, the mass splits of the heavy Majorana neutrinos
$N_{i}^{\pm}$ can be determined by
\begin{eqnarray}
&&r_{N_i^{}}^{}=\mathcal{O}\left(10^{-12}_{}\right)~~\textrm{for}~~\nonumber\\
&&h_{\xi_{11}^{}}^{}=\mathcal{O}\left(10^{-5}_{}\right)\,,~~
h_{\xi_{22,33}^{}}^{}=\mathcal{O}\left(10^{-4}_{}\right)\,.
\end{eqnarray}
The sterile neutrinos $S_i^{}$ also obtain small masses:
\begin{eqnarray}
m_{s_i^{}}^{}=\mathcal{O}(1-100\,\textrm{keV}) ~~\textrm{for}~~
h_{\zeta_{i}^{}}^{}=\mathcal{O}\left(10^{-4}_{}-10^{-2}_{}\right)\,.
\end{eqnarray}
Thirdly, we consider
\begin{eqnarray}
-[\mathcal{O}(10^2_{}\,\textrm{GeV})]=\mu_\varphi^2<0<\mu_\eta^2=[\mathcal{O}(10^5_{}\,\textrm{GeV})]^2_{}\,,
\end{eqnarray}
to derive the mixing angle:
\begin{eqnarray}
\vartheta\simeq -\frac{\kappa v_\chi^{}
v_\sigma^{}}{\mu_\eta^2}=10^{-4}_{}\quad \textrm{for}\quad
\kappa=\mathcal{O}(1)
\end{eqnarray}
and then determine the VEVs:
\begin{eqnarray}
v_\varphi^{}\simeq 174\,\textrm{GeV}\,,~~
v_\eta^{}\simeq-\frac{\kappa v_\chi^{} v_\sigma^{}
v_\varphi^{}}{\mu_\eta^2}\simeq v_\varphi^{}\vartheta\,.
\end{eqnarray}

We now consider the following sample of the Yukawa couplings,
\begin{eqnarray}
\label{parameter1} \left|y_{\nu_{e 1}^{}}^{}\right|=1.15\times
10^{-6}_{}\,,\,\left|y_{\xi_{e
1}^{}}^{}\right|=10^{-3}\,,\,\,\,\sin\delta_{e
1}^{}=~~\,1\,,&&\nonumber\\
\left|y_{\nu_{\mu 1}^{}}^{}\right|= 1.15\times
10^{-7}_{}\,,\,\left|y_{\xi_{\mu
1}^{}}^{}\right|=\frac{10^{-2}_{}}{2}\,,\,\sin\delta_{\mu
1}^{}=-1\,,&&\nonumber\\
\left|y_{\nu_{\tau 1}^{}}^{}\right|=1.15\times
10^{-7}_{}\,,\,\left|y_{\xi_{\tau 1}^{}}^{}\right|=\frac{
10^{-2}_{}}{2}\,,\,\sin\delta_{\mu 1}^{}=-1\,,&&
\end{eqnarray}
to fulfill the assumption (\ref{cpassumption}). We then read
\begin{eqnarray}
K_{N_1^+}^{}=K_{N_1^-}^{}=0.135\,.
\end{eqnarray}
By further fixing $r_{N_1^{}}^{}=10^{-12}$, we can obtain the CP
asymmetry,
\begin{eqnarray}
\varepsilon_{N_1^+}^{\nu_{e}^{}}=\varepsilon_{N_1^-}^{\nu_{e}^{}}=2.03\times
10^{-3}_{}\,.
\end{eqnarray}
In consequence, the large lepton asymmetries should be
\begin{eqnarray}
\eta_{\nu_e^{}}^{}=3.63\times
10^{-5}_{}=\frac{6.99\times10^{-5}_{}}{1+4\sin^2\theta_W^{}}\,,
\end{eqnarray}
which is in agreement with the observations (\ref{bbn}). Accordingly
a desired baryon asymmetry is induced by
\begin{eqnarray}
\eta_B^{}&\simeq &\frac{6}{13\pi^2_{}}\frac{f_\mu^2+xf_\tau^2}{1+x}
\eta_{\nu_e^{}}^{}=0.888\times 10^{-10}_{}~~\textrm{for}\nonumber\\
&&x=\frac{\left|y_{\nu_{\tau 1}^{}}^{}\right|\left|y_{\xi_{\tau
1}^{}}^{}\right|\sin\delta_{\tau 1}^{}}{\left|y_{\nu_{\mu
1}^{}}^{}\right|\left|y_{\xi_{\mu 1}^{}}^{}\right|\sin\delta_{\mu
1}^{}}=1\,.
\end{eqnarray}
We further take
\begin{eqnarray}
\label{parameter2} y_{\nu_{\alpha 2}^{}}^{}= y_{\nu_{\alpha
3}^{}}=100\,y_{\nu_{\alpha 1}^{}}^{}\,,~~\mu_{1j}^{}=
20\,\textrm{keV}\,,
\end{eqnarray}
and then perform
\begin{eqnarray}
4\theta_{e 1}^2= 10^{-12}_{}\,,~~4\theta_{\mu 1}^2=4\theta_{\tau
1}^2=10^{-14}_{}\,.
\end{eqnarray}
So the $\nu_{e}^{}\rightarrow S_1^{}$ oscillation is dominant.
Compared with the fitting by \cite{blrv2009-2,brs2009}, we see the
sterile neutrino $S_1^{}$ with the mass
\begin{eqnarray}
m_{S_1^{}}^{}\simeq 8\,\textrm{keV}
\end{eqnarray}
can be a good candidate for the dark matter. The other sterile
neutrinos $S_{2,3}^{}$ are also at the keV scale. They could leave a
significant relic density if their mixing with the active neutrinos
$\theta_{\alpha i}^2~(i=2,3)$ are comparable with $\theta_{\alpha
1}^2 $. Alternatively, their relic density could be negligible if
$\theta_{\alpha i}^2~(i=2,3)$ are much smaller than $\theta_{\alpha
1}^2$. This could be achieved by taking appropriate
$\mu_{ij}^{}(i=2,3)$ in Eq. (\ref{asmixing}).

For the above parameter choice, the seesaw contribution from the
sterile neutrinos to the neutrino masses is of the order of
$\theta_{\alpha i}^2
m_{S_i^{}}^{}\lesssim\mathcal{O}(10^{-9}_{}\,\textrm{eV})$, which is
too small to explain the neutrino oscillation data. Furthermore, the
contribution from the heavy Majorana neutrinos $N_1^\pm$ is also too
small ($\sim\mathcal{O}(10^{-5}_{}\,\textrm{eV})$). Fortunately, the
other heavy Majorana neutrinos $N_{2,3}^\pm$ can give a desired
contribution. Specifically, we can take the Yukawa couplings
$y_{\xi_{\alpha i}^{}}^{}\lesssim \mathcal{O}(1) (i=2,3)$ in the
neutrino mass matrix which is dominated by the second term of Eq.
(\ref{numass1}). We also check the lepton number violating processes
mediated by $N_{2,3}^{\pm}$ have been decoupled before the
leptogenesis epoch since their reaction rate \cite{fy1990},
\begin{eqnarray}
\Gamma_A^{}=\frac{1}{\pi^{3}_{}}\frac{\textrm{Tr}(m_\nu^\dagger
m_\nu^{})}{v^4_{}}T^{3}_{}=\frac{1}{\pi^{3}_{}}\frac{\sum_{i}^{}
m_i^2}{v^4_{}}T^{3}_{}\,,
\end{eqnarray}
is smaller than the Hubble constant at the temperature $T=
M_{N_1^\pm}^{}$.

\section{summary}

We have shown the resonant leptogenesis in the seesaw context for
the observed small baryon asymmetry can allow large neutrino
asymmetries of individual flavors for the production of the sterile
neutrino dark matter due to the thermal mass effects of the
sphaleron processes. In our model, the inflaton can nicely obtain a
small VEV as a result of the seesaw suppressed ratio of the $B-L$
breaking scale over its heavy mass. The quasi-degenerate mass
spectrum of the heavy Majorana neutrinos for the resonant
leptogenesis is naturally given by the scale of the $B-L$ symmetry
breaking and the small VEV of the inflaton. The sterile neutrinos
obtain small Majorana masses through their Yukawa couplings with the
inflaton. Their mixing with the active neutrinos are suppressed by
the seesaw mechanism. Thus one or more sterile neutrinos can act as
the warm dark matter in the presence of the resonant active-sterile
neutrino oscillation. Due to the small active-sterile mixing, the
sterile neutrinos only have a negligible contribution to the
neutrino masses. Instead, the seesaw contribution from the heavy
Majorana neutrinos is responsible for generating the desired
neutrino masses.

For an appropriate parameter choice, we may realize the $\nu$MSM
model \cite{als2007,as2005,brs2009} with a pair of
quasi-degenerately heavy Majorana neutrinos and a sterile neutrino
in our model with three pairs of quasi-degenerately heavy Majorana
neutrinos and three sterile neutrinos. In the $\nu$MSM model, the
processes for generating the lepton asymmetry can keep working below
the sphaleron freeze-out temperature at which there is no conversion
of the lepton asymmetry to the baryon asymmetry. In this case, we
need not resort to the thermal mass effects of the sphaleron
processes.

\vspace{5mm}

\textbf{Acknowledgement}: I thank Manfred Lindner for hospitality at
Max-Planck-Institut f\"{u}r Kernphysik. This work is supported by
the Alexander von Humboldt Foundation.

\end{document}